\documentclass[aps,twocolumn,prl,superscriptaddress,showpacs,showkeys,floatfix]{revtex4-1}
\usepackage{graphicx,color}
\usepackage{float}
\usepackage[caption = false]{subfig}
\usepackage{epsfig}
\usepackage{graphicx}
\usepackage{mathrsfs}
\usepackage{color}
\usepackage{bm}
\usepackage{amsmath,amssymb,amsthm,amscd}
\usepackage{bbold}
\usepackage{mdwlist}

 \newcommand{\ket}[1]{|{#1}\rangle  }
 \newcommand{\braket}[2]{\langle{#1}| {#2}\rangle}
 \newcommand{\ketbra}[2]{\vert {#1} \rangle \langle{#2}\vert}

\newcommand{\Tr}{\operatorname{Tr}}

\allowdisplaybreaks

\begin{document}

\title{Efficient accessible bounds to the classical capacity of quantum channels}

\author{Chiara Macchiavello}
\affiliation{Quit group, Dipartimento di Fisica, 
Universit\`a di Pavia, via A. Bassi 6, 
 I-27100 Pavia, Italy}
\affiliation{Istituto Nazionale di Fisica Nucleare, Gruppo IV, via A. Bassi 6,
  I-27100 Pavia, Italy}

\author{Massimiliano F. Sacchi}
\affiliation{Istituto di Fotonica e Nanotecnologie - CNR, Piazza Leonardo
  da Vinci 32, I-20133, Milano, Italy}
\affiliation{Quit group, Dipartimento di Fisica, 
Universit\`a di Pavia, via A. Bassi 6, 
 I-27100 Pavia, Italy}
	
\date{\today}

\begin{abstract} 
We present a method to detect lower bounds to the classical capacity
of quantum communication channels by means of few local measurements
(i.e. without complete process tomography), reconstruction of sets of
conditional probabilities, and classical optimisation.  The method
does not require any a priori information about the channel. We illustrate
its performance for significant forms of noisy channels.
\end{abstract}

\maketitle

The classical capacity of a noisy quantum communication channel
quantifies the maximum amount of classical information per channel use
that can be reliably transmitted \cite{NC00}.  In general, its
computation is a hard task, since it requires a regularisation
procedure over an infinite number of channel uses \cite{hol0,sw,hol},
and it is therefore by itself not directly accessible
experimentally. Its analytical value is known mainly for some channels
that have the property of additivity, since regularisation is not
needed in this case. In fact, in such case the problem is recast to
the evaluation of the Holevo capacity \cite{hol0,sw,hol}, which is a
single-letter expression quantifying the maximum information when only
product states are sent through the uses of the channel.

When a complete knowledge of the channel is available, then several
methods can be used to calculate the Holevo capacity
\cite{eff,eff2,eff3,eff4,eff5}, which is always a lower bound to the
ultimate capacity of the channel.  In many practical situations,
however, a complete knowledge of the kind of noise present along the
channel is not available, and sometimes noise can be completely
unknown.  It is then important to develop efficient means to establish
whether in these situations the channel can still be profitably
employed for information transmission.  A standard method to establish
this relies on quantum process tomography
\cite{nielsen97,pcz,mls,dlp,alt,cnot,vibr, ion,mohseni,irene,atom},
where a complete reconstruction of the completely positive map
describing the action of the channel can be achieved, but it is a
demanding procedure in terms of the needed number of different
measurement settings, since it scales as $d^4$ for a finite
$d$-dimensional quantum system.

\par When one is not interested in reconstructing the complete form of
the noise but only in detecting lower bounds to the classical capacity
a novel and less demanding procedure in terms of measurements is
presented in this Letter. In the same spirit as it is done, for
example, for detection of entanglement-breaking property
\cite{qchanndet,qchanndet2} or non-Markovianity \cite{nomadet} of
quantum channels, or for detection of lower bounds to the quantum
capacity \cite{ms16}, the method we present allows to experimentally
detect lower bounds to the classical capacity by means of a number of
local measurements that scales at most as $d^2$.  The method proposed in
Ref. \cite{ms16} for detecting bounds $Q_{DET}$ to the quantum
capacity $Q$ can be applied to generally unknown noisy channels and
has been proved to be very successful for many examples of single
qubit channels, for generalized Pauli channels in arbitrary dimension,
and for two-qubit memory Pauli and amplitude damping channels
\cite{ms-corr}.  Moreover, the first experimental demonstration has
been also recently shown in Ref. \cite{exp}, based on a quantum
optical implementation for various forms of noisy channels.

In principle, since the procedure of Ref. \cite{ms16} allows to
experimentally reconstruct bounds $Q_{DET}$ on the quantum capacity
$Q$, private capacity $P$, and entangled-assisted capacity $C_{EA}$,
and since for the classical capacity $C$ generally one has
\cite{mixed,chain} the chain of inequalities $C \geq C_1 \geq
C_{EA}-\log _2 d \geq Q_{DET}$, the method to measure $Q_{DET}$ also
gives a bound to the classical capacity. However, unfortunately, such
a bound is in general very loose.  In the present Letter we will show
a more effective method constructed specifically to measure a lower
bound to the classical capacity.

Quantum channels are described by completely positive and trace
preserving maps ${\cal E}$, which can be expressed in the Kraus form
\cite{NC00} as ${\cal E}(\rho )=\sum_k A_k\rho A_k^\dagger $, where
$\rho$ is the density operator of the quantum system on which the
channel acts and the Kraus operators $\{A_k\}$ fulfill the constraint
$\sum_k A_k^\dagger A_k= I$. The classical capacity $C$ of the channel
${\cal E}$ quantifies the maximum number of bits per channel use that
can be reliably transmitted through the noisy quantum channel. It is
defined \cite{hol0,sw,hol} by the regularized expression $C=\lim
_{n\rightarrow \infty} \chi ({\cal E}^{\otimes n})/n$, in terms of the
Holevo capacity
\begin{eqnarray}
\chi (\Phi) = \max _{\{p_i , \rho _i \}}S[\Phi (\textstyle \sum _i p_i \rho
  _i)]-\textstyle 
\sum
_i p_i S[\Phi (\rho _i)]
\;,
\end{eqnarray}
where the maximum is over all possible ensembles of quantum states,
and $S(\rho )=-\Tr[\rho \log \rho ]$ denotes the von Neumann entropy
(we use logarithm to the base $2$). The Holevo capacity $\chi ({\cal
  E})\equiv C_1$ is a lower bound for the channel capacity, and
corresponds to the maximum information when only product states are
sent through the uses of the channel, whereas joint (entangled)
measurements are allowed at the output. Then, clearly, the Holevo
capacity is also an upper bound for any expression of the mutual
information \cite{mutu,mutu2,mutu3}
\begin{eqnarray}
I(X;Y)=\sum _{x,y}p_x p(y|x) \log \frac {p(y|x)}{\sum _{x'} p_{x'} p(y
  |x')}\;,
\end{eqnarray}
where the probability transition matrix $p(y|x)$ corresponds to the
conditional probability of an arbitrary measurement with outcome $y$
at the output for a single use of the channel with input $\rho _x$,
and $p_x$ denotes an arbitrary prior probability, which describes the
distribution of the encoded alphabet.
 
Here, we are interested to detect a lower bound to the capacity when
the number of measurement settings is smaller than the one needed for
complete process-tomography, going from a single one to $d^2 -1$.   
The scenario we will focus on to achieve
our detection strategy consists in the following steps: prepare a
bipartite maximally entangled state $|\phi^+\rangle =\frac {1}{\sqrt
  d} \sum_{k=0}^{d-1} |k \rangle |k \rangle $ of a system qudit and a
noiseless reference qudit, and send it through the channel ${\cal
  E}\otimes{\cal I} _R$, where the unknown channel ${\cal E}$ acts on
the system qudit alone. Then measure locally a number of observables of
the form $X_i\otimes X_i ^{T}$, where $T$ represents the transposition
w.r.t. to the fixed basis defined by $|\phi ^+ \rangle $.

By denoting the $d$ eigenvectors of $X_i$ as $\{|\phi ^{(i)}_n \rangle
\}$ and using the identity \cite{pla}
\begin{equation}
\Tr [(A\otimes B^T)({\cal E}\otimes{\cal I} _R)
|\phi ^+ \rangle \langle \phi ^+ |]=\frac 1d   
\Tr[A {\cal E}(B)]\;,
\end{equation} 
it is straightforward to see that the measurement protocol allows us
to reconstruct the conditional probabilities $p^{(i)}(m|n) =
\braket {\phi ^{(i)}_m} { {\cal E} (\ketbra {\phi ^{(i)}_n} {\phi
    ^{(i)}_n}) | {\phi ^{(i)}_m} }$.  We can then write the optimal
mutual information for the encoding-decoding scheme $(i)$ as
\begin{eqnarray}
I^{(i)}=\max _ {\{p_n^{(i)} \}} \sum _{n,m}p_n^{(i)} 
p^{(i)}(m|n)  
\log \frac{p^{(i)}(m|n)  }{\sum _l 
p_l^{(i)} p^{(i)}(m|l) }
\;.\label{ii}
\end{eqnarray}
Then, one has  the following chain of inequalities
\begin{eqnarray}
C ({\cal E})
\geq C_1 ({\cal E}) \geq C_{DET} \equiv \max _i \{I^{(i)}\}
\;,
\end{eqnarray}
where $C_{DET}$ is the experimentally accessible bound to the
classical capacity, which depends on the chosen set of measured
observables labeled by $i$. It is then clear that the bound improves
by increasing the number of performed measurements. 

We want to point out that such a detection method based on the
measurements of the local operators does not necessarily require the
use of an entangled bipartite state at the input. Actually, each
conditional probability $p^{(i)}(m|n)$ can be obtained equivalently by
considering only the system qudit, preparing it in the eigenstates of
$X_i$ with equal probabilities, and measuring $X_i$ at the output of
the channel. We notice also that if one has the possibility of
optimizing over all input ensembles, whereas the output measurements
are fixed, the problem is recast to evaluate the informational power
of the noisy quantum measurements \cite{dalla} corresponding to the
Heisenberg evolution ${\cal E}^\vee (\{ |\phi ^{(i)}_n \rangle \langle
\phi ^{(i)}_n |\})$, and then maximise over $i$.

The maximisation over the set of prior probabilities $\{p_n ^{(i)} \}$
in Eq. (\ref{ii}) for each $i$ can be achieved by means of the
Blahut-Arimoto recursive algorithm \cite{bga1,bga2,bga3}, given by
\begin{eqnarray}
&&c_n ^{(i)}[r]
=\exp \left (\sum _m p^{(i)}(m|n)  \log \frac{p^{(i)}(m|n)  }{\sum _l 
p_l^{(i)}[r] p^{(i)}(m|l) } \right ) \,;
\nonumber \\& & 
p_n ^{(i)}[r+1]=p_n ^{(i)}[r] \frac {c_n ^{(i)}[r]}
{\sum _l p_l ^{(i)}[r] c_l ^{(i)}[r]} 
\;.\label{ari}
\end{eqnarray}
Starting from an arbitrary prior $\{p_n ^{(i)} [0]\}$, this guarantees
convergence to an optimal prior, thus providing the value of $I^{(i)}$
for each $i$ with the desired accuracy.  For completeness we mention
that a slight modification of the recursive algorithm (\ref{ari}) can
also accommodate possible constraints of the channel, e.g. the allowed
maximum energy in lossy Bosonic channels \cite{loss}.

\par We remind that in some special forms of transition matrices
$p^{(i)}(m |n)$ there is no need of numerical maximisation, since the
optimal prior is known. This is the case of a conditional probability
$p^{(i)}(m|n)$ corresponding to a weakly symmetric channel \cite{CT},
where every column $p^{(i)}(\cdot |n)$ is a permutation of each other
and all sums $\sum_{n} p^{(i)}(m |n) $ are equal. In fact, in such
case the corresponding optimal prior is the uniform $p^{(i)}_n=1/d$,
and the mutual information is given by $I^{(i)}=\log d -
H[p^{(i)}(\cdot | n)]$, where $H(\vec x)=-\sum _i x_i \log x_i$
denotes the Shannon entropy and therefore $H[p^{(i)}(\cdot | n)]$ is
the Shannon entropy of an arbitrary column (since all columns have the
same entropy).
\par A further example is the case of the transition matrices obtained
by preparation and measurements on the eigenstates of the generalized
Pauli (or Weyl) matrices $U_{ls}=\sum _{k=0}^{d-1} e^{\frac{2\pi i}{d}
  kl} |k \rangle \langle (k + s)\!\!\!\mod d |$, with $l,s\in
[0,d-1]$, and the channel is a generalized Pauli channel ${\cal
  E}(\rho )=\sum_{l,s=0}^{d-1} q_{ls} U_{ls} \rho U^ \dag _{ls}$
\cite{bellobs}. In fact, when $d$ is prime, for each $(l,s)$ all
columns (and rows) of $p^{(ls)}(m|n)$ are permutations of each other
\cite{weyl}, i.e. one has a symmetric channel, and then each optimal
prior is uniform \cite{CT}.  In this case, then, the detected capacity
will be given by
\begin{eqnarray}
C_{DET}=\log d -  \min _{\{l ,s \}} H[p^{(ls)}(\cdot |n)]
\;.\label{weyl}
\end{eqnarray}
Actually, the present result has been recently used to provide
efficient detectable bounds to the Holevo capacity of generalized
Pauli channels \cite{weyl2}.  
\par As a final example we mention the
case of binary channels, where $p^{(i)}(m|n)$ are $2\times 2$
transition matrices, and the maximisation in Eq. (\ref{ii}) can be
analytically found (see the Supplemental Material).

In order to study in more detail the case of qubit channels, we use
their representation on the
Bloch sphere. A qubit channel
$\Phi$ maps the sphere of possible input states to an ellipsoid, and may be
expressed as \cite{ellip,ellip2}
\begin{equation}
\Phi(\rho) = \frac 12 [I + \vec \sigma \cdot (\mathbf{\Lambda} \vec r
  + \vec t)  ], \label{aff}
\end{equation}
where $\vec r$ denotes the Bloch vector of the input state $\rho $,
$\mathbf{\Lambda }$ is a real $3\times 3$ matrix, 
and $\vec t$ is the output Bloch vector for input $\rho =\frac I2$. 
That is, the channel $\Phi$ produces the affine mapping $\vec r \mapsto
\mathbf{\Lambda} \vec r + \vec t$. Via local unitary operations acting 
before and after the map, the transformation matrices $\mathbf{\Lambda}$ and
$\vec t$ may be brought to the form
\begin{equation}
\label{em}
\mathbf{\Lambda} = \left( \begin{array}{*{20}c}
   \lambda_1 & 0 & 0 \\
   0 & \lambda_2 & 0 \\
   0 & 0 & \lambda_3 \\
\end{array}
\right), \qquad
\vec t = \left( \begin{array}{*{20}c}
   t_1 \\
   t_2 \\
   t_3 \\
\end{array}
\right),
\end{equation}
namely, an arbitrary qubit channel $\Phi$ may be expressed as $\Phi = \Gamma_U
\circ \Phi_{\vec t,\mathbf{\Lambda}} \circ \Gamma_V$, where $\Gamma_U$ and $\Gamma_V$ are
unitary channels (thus not varying the capacity), 
and $\Phi_{\vec t, \mathbf{\Lambda}}$ is the channel with $\mathbf{\Lambda}$
and $\vec t$ given by Eq. \eqref{em}. So the Bloch sphere is mapped to an
ellipsoid with principal axes parallel to the Cartesian axes, and
center shifted by the vector $\vec t$.  We recall that the condition of
complete positivity puts some constraints on the admissible values of
$\lambda _i$'s and $t_i$'s \cite{algoe,ellip,ellip2,braun}. 

\par Let us consider our detection scheme for channels of the form $\Phi
_{\vec t, \mathbf{\Lambda}}$ and measurements of the Pauli operators.    
The conditional probabilities of  outcomes $\pm 1$ for the measurement of
$\sigma _{i}$ at the output of a channel $\Phi_{\vec t,\mathbf{\Lambda}}$ 
given the input states 
$\rho ^{(i)}_{+1}=\frac 12 (1+ \sigma _i)$ 
and $\rho ^{(i)}_{-1}=\frac 12 (1- \sigma _i)$ are given respectively by  
\begin{eqnarray}
&&p^{(i)}(\pm 1 |+ 1)= 
\frac 12 [1\pm (t_i +\lambda _i )] \,,\nonumber \\& & 
p^{(i)}(\pm 1 | - 1)= 
\frac 12 [1\pm (t_i- \lambda _i)] 
\;.
\end{eqnarray}
Upon defining $\epsilon _0^{(i)}= \frac 12 (1-\max \{ |\lambda_i +t_i
|,|\lambda _i -t_i|\})= \frac 12 (1- |\lambda _i|-|t_i|)$ 
and $\epsilon _1 ^{(i)}=\epsilon _0 ^{(i)}+ |t_i|$, 
each of the three measurements then provides a transition matrix of a
binary channel, and the detected bound for the classical capacity takes the form
\begin{eqnarray}
C_{DET}= \max _i \{C_B(\epsilon _0 ^{(i)},\epsilon _1 ^{(i)})\}
\label{form}\;,
\end{eqnarray}
with $C_B(\epsilon _0, \epsilon _1)$ given by Eq. (A.4) in the
Supplemental Material.

The case of Pauli channels ${\cal E}(\rho )=(1-p_x-p_y-p_z)\rho +
p_x \sigma _x \rho \sigma _x + p_y \sigma _y \rho \sigma _y +p_z \sigma _z \rho \sigma _z
$ corresponds to the unital case, i.e. ${\cal
  E}(I)=I$, for which $\lambda _1=1-2p_y-2p_z$,
$\lambda _2=1-2p_x-2p_z$, $\lambda _3=1-2p_x-2p_y$, and
$t_1=t_2=t_3=0$. Hence, the detected capacity in this case is given by 
\begin{eqnarray}
&&C_{DET}=1-H\left(\frac {1 -\max _i \{|\lambda _i|\}}{2} \right
  )= \label{unc} \\& & 1-
\min \{H(p_y+p_z),H(p_x+p_z),H(p_x+p_y) \}\;,\nonumber 
\end{eqnarray}
which equals the Holevo and the classical capacity (since the
additivity hypothesis holds true for unital qubit channels
\cite{unital}), 
i.e. $C=C_1=C_{DET}$, and just corresponds to the result of
Eq. (\ref{weyl}) for $d=2$.    

It is useful to recall the following results from \cite{pseudo}, where
the notion of pseudoclassical channel is introduced.  In a
pseudoclassical channel the Holevo capacity $C_1$ is achieved without
quantum correlations in the measurements between different uses of the
channel. In other words, for pseudoclassical channels the Holevo
capacity can be attained when the optimal measurement with a single
transmission is performed on each system. A channel is pseudoclassical
iff the capacity $C_1$ is achieved by an ensemble of input states such
that the corresponding output states are mutually commuting.  All
unital qubit channels are pseudoclassical \cite{pseudo} and, as we
have seen, $C=C_1=C_{DET}$. In the nonunital case, a channel
(\ref{aff}) is pseudoclassical iff \cite{pseudo,hay}

i) $\vec t \neq 0$ is an eigenvector of $\mathbf{\Lambda }\mathbf{\Lambda }^\dag $ with
eigenvalue $r^2$, and 

ii) by denoting as $r_0 ^2$ the maximum of the other two eigenvalues,  
one has 
\begin{eqnarray}
&&r_0^2\leq T(\|\vec t \|, r) \equiv r^2 - \| \vec t \| r + \nonumber
  \\& &
 \frac{
(\|\vec t\| -r)\left [ 
H\left(\frac{1+\|\vec t\|+r}{2}\right )-H\left(\frac{1+\|\vec
    t\|-r}{2}\right )
\right] } { H'\left(\frac{1+\|\vec t\|-r}{2} \right )}
\;,\label{13}
\end{eqnarray}
where $H'=\frac{d H(x)}{dx}=\log \frac{1-x}{x}$.  
Geometrically, condition i) requires that the direction of the shift must be
parallel to one of the principal axes, and  condition ii) requires that the
ellipsoid must be sufficiently thin around this direction. 
For  qubit nonunital pseudoclassical channels the Holevo capacity is given by \cite{pseudo}
\begin{eqnarray}
C_1=C_B \left ( \frac { 1- \|\vec t \| -r }{2},
\frac {1 +\|\vec t \| -r }{2} \right)
\;,
\end{eqnarray} 
namely it corresponds to the capacity of a classical binary asymmetric
channel. 

In the following we consider the case of Eq. (\ref{em}) with $t_1=t_2=0$ and  $t_3=t$,
which are the most studied non-unital channels in the literature
\cite{eff3,nata,berry,daems}. 
In this case the
condition of complete positivity is equivalent \cite{algoe,ellip,ellip2} to the constraints 
$(\lambda _1 \pm \lambda _2)^2 \leq (1\pm \lambda _3)^2 -t^2 $. 
Note that the condition of pseudoclassicality (\ref{13}) rewrites as 
\begin{eqnarray}
\lambda _m ^2 \equiv \max\{\lambda _1 ^2, \lambda _2 ^2 \}\leq T (t , \lambda _3)\,.
\label{lmm}
\end{eqnarray}
When this condition is satisfied the
detected capacity provides the exact value of the Holevo capacity, namely  
\begin{eqnarray}
C_1 = C_{DET}= C_B (\epsilon _0 ^{(3)},\epsilon _1 ^{(3)}) \;,\label{lmm2}
\end{eqnarray}
which is achieved by  
the two orthogonal eigenstates of $\sigma _z$ as input states. 
This is also due to the fact that the corresponding output states 
$\rho =\frac 12 [I \pm (\lambda _3 +t_3)\sigma _z]$ 
are commuting, and hence the Holevo information
is saturated by the projective measurement on $\sigma _z$ eigenstates
\cite{fuchs}. 
On the other hand, when condition (\ref{lmm}) is not  satisfied, one has $C_{DET}<
C_1$, since typically the capacity $C_1$ is achieved by an ensemble of
two nonorthogonal input states, or even a three- or four-state
ensemble \cite{nata,eff3}, while only input orthogonal states are
tested  by our detection
method. In the following we test our method on further explicit
examples.  

\begin{figure}[htb]
  \includegraphics[scale=0.42]{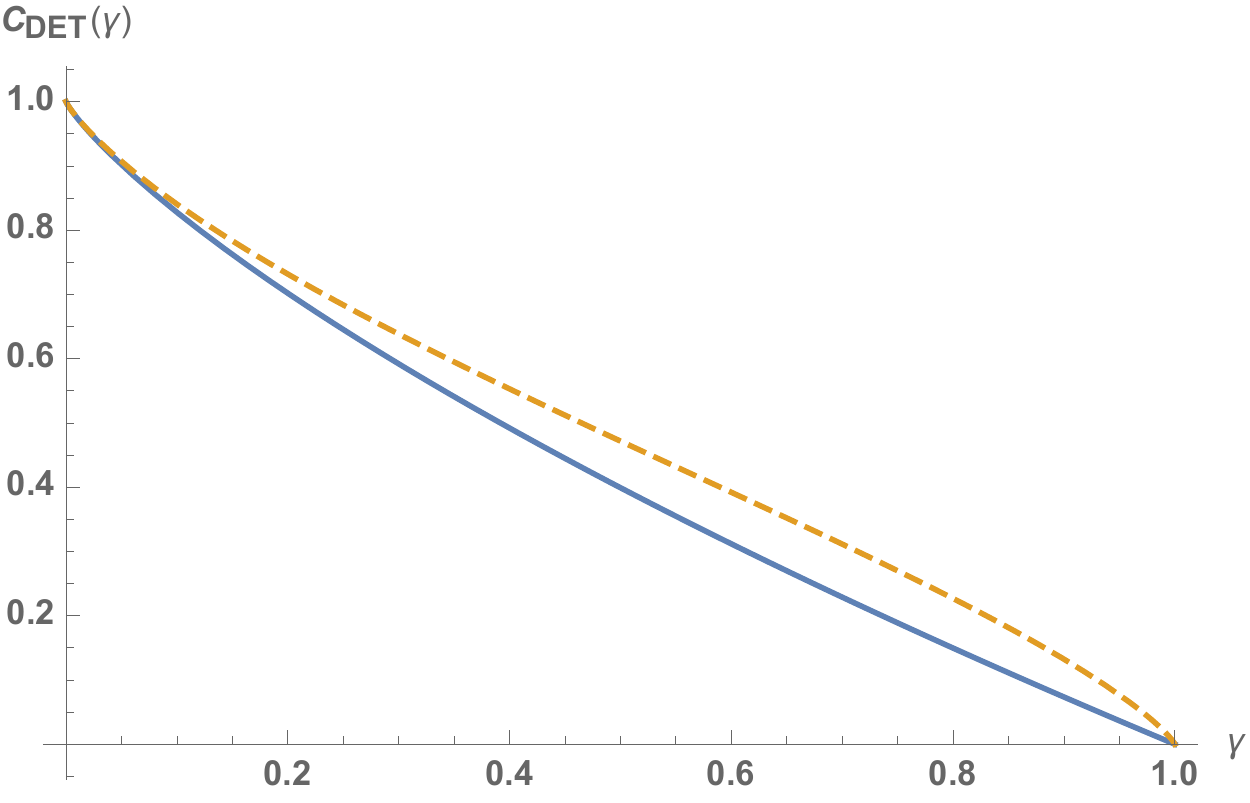}
  \caption{Detected classical capacity $C_{DET}(\gamma )$ for a qubit generalized
    damping channel versus damping parameter (solid line), along with the 
    Holevo capacity $C_1(\gamma ,1)$ for the damping channel
    with $p=1$ (dotted line).}
  \label{fig:det1}
\end{figure}

Example: generalized amplitude damping channel, for which $\lambda _1 =\lambda
_2 =\sqrt{1-\gamma}$, $\lambda _3 =(1-\gamma )$, and $t_3
=(2p-1)\gamma$, where both $\gamma $  and $p \in [0,1]$.  
This channel describes qubit dynamics with exchanges of excitations
with the thermal environment at finite temperature \cite{NC00,turch,turch2}.   
Then one has $\epsilon _0^{x}=\epsilon _1^{x}=\epsilon _0^{y}=\epsilon _1^{y}=
(1-\sqrt{1-\gamma })/2 $, and $\epsilon _0^{z}=\gamma \min \{p,1-p\}$, 
$\epsilon _1 ^{z}=\gamma \max \{p,1-p \}$. It follows that the detected
classical capacity is given by 
\begin{eqnarray}
&&C_{DET}(\gamma , p)= \max\Big \{ 
1- H\Big ( \frac{1-\sqrt{1-\gamma
}}{2}\Big ), \nonumber \\& & 
\ C_B (\gamma \min \{p,1-p \}, \gamma \max \{p,1-p \})\Big \}\;.\label{cgp1}
\end{eqnarray}
We numerically checked that the maximum is always achieved by the
first term in Eq. (\ref{cgp1}), which is independent of the value of
$p$. 
In fact, one can also check that condition (\ref{lmm}) is never
satisfied, and so the channel is never pseudoclassical (except the
degenerate case $p=\frac 12$, when the channel becomes unital). Hence, the
detected capacity is strictly lower than the
Holevo capacity. For $p=1$, corresponding to the customary amplitude damping
channel at zero temperature, one can compute the Holevo capacity according to 
the following equation \cite{gf,gf2}
\begin{eqnarray}
C_1(\gamma ,1)= \max _{t\in [0,1]} \left (
 H[t (1 - \gamma ) ] - H\left [g(\gamma ,t)
\right ]\right )
\;,\label{c1g}
\end{eqnarray}
with $g(\gamma ,t)\equiv\frac{1 + \sqrt{1 - 4 \gamma (1 - \gamma )
    t^2}}{2}$.  Clearly, $C_1(\gamma ,1)$ provides a known lower bound
to the classical capacity.  In Fig.  \ref{fig:det1} we plot the
detected capacity $C_{DET}(\gamma )\equiv C_{DET}(\gamma , p)$ versus
the damping parameter $\gamma $, along with the Holevo capacity
$C_1(\gamma ,1)$.  Notice that $C_{DET}(\gamma
)=C_1 (\gamma, 1/2)$, which corresponds to the capacity of a Pauli
channel.

Further examples corresponding to a stretched damping channel
\cite{nata} and to extremal qubit channels \cite{ellip2,braun} are
reported in the Supplemental Material.

\begin{figure}[htb]
  \includegraphics[scale=0.46]{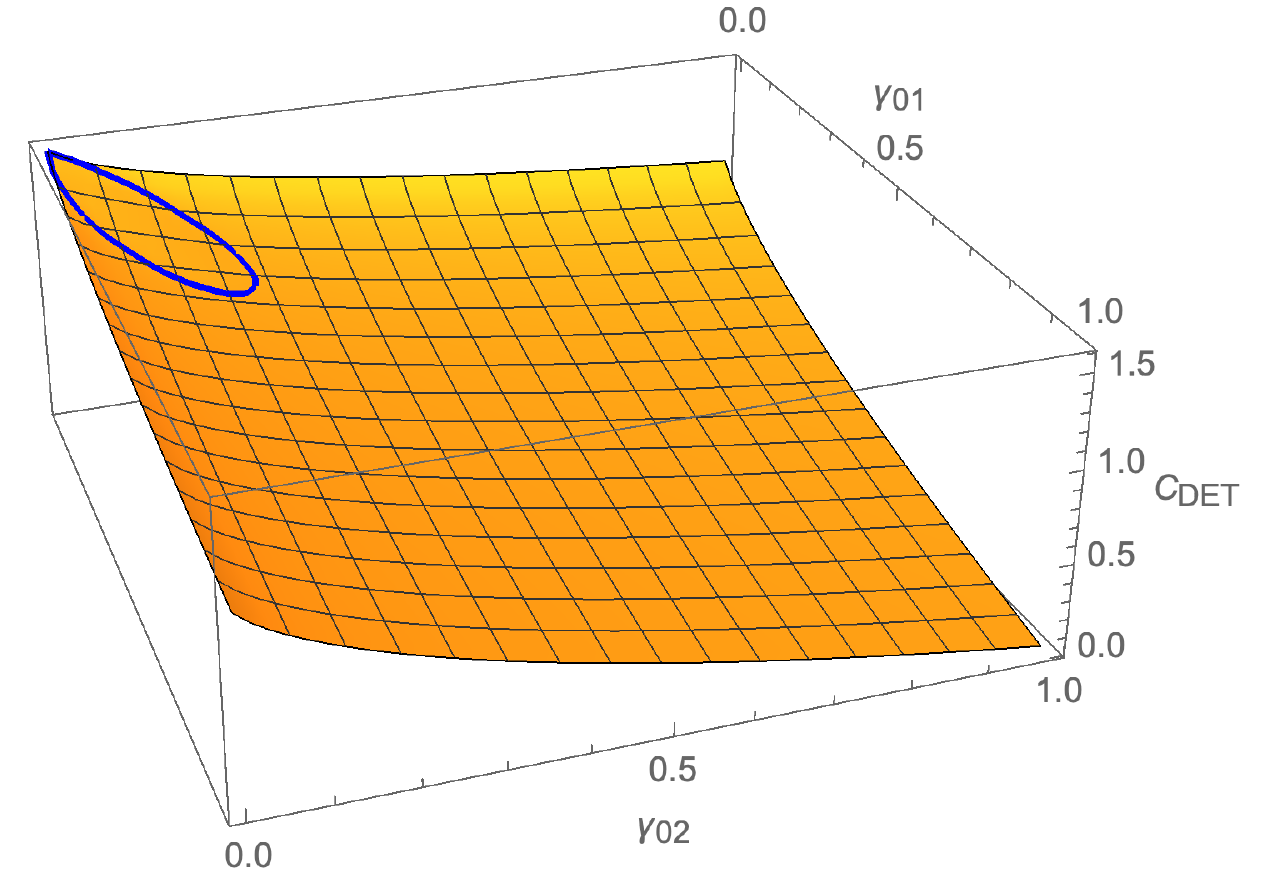}
  \caption{Detected classical capacity $C_{DET}(\gamma _{01},\gamma
    _{02})$ 
for a qutrit damping channel with $V$-configuration versus damping
parameters $\gamma _{01}$ and $\gamma _{02}$. Inside (outside) the enclosed
region, the detected capacity is achieved by the basis $B_2$  ($B_1$) in Eq. (\ref{bunodue}).}  
  \label{fig:c3damp}
\end{figure}

Our method can also be successfully applied to other forms of amplitude damping
processes. For instance, when $d=3$, we can consider 
the decay processes of a three-level system in $V$-shaped
configuration \cite{qutrit}. The Kraus operators take the form
\begin{eqnarray}
&&A_0{}={\rm diag}\{1,\sqrt{1-\gamma_{01}},\sqrt{1-\gamma_{02}}\},\nonumber\\
&&A_1{}=\sqrt{\gamma_{01}}|0\rangle\langle1|,\;\;A_2{}=\sqrt{\gamma_{02}}|0\rangle\langle2|,
\;\label{kra3}
\end{eqnarray}
with both $\gamma _{01}$ and $\gamma _{02} \in [0,1]$. 
We consider the case where only two projective measurements are used
to detect the classical capacity, namely the two mutually unbiased bases 
\begin{eqnarray}
&&B_1 =\left \{ |n \rangle _1 =|n \rangle ,\  n=0,1,2 \right \} \,;\nonumber \\& & 
B_2=\left \{|n \rangle _2 =
(\textstyle \sum _{j=0}^2 \omega ^{nj} |j \rangle )/{\sqrt 3},\ 
n=0,1,2 \right \}\,,\label{bunodue}
\end{eqnarray}
with $\omega =e^{2\pi i/3}$. The corresponding transition matrices 
$Q^{(1)}$ and $Q^{(2)}$  are reported
in the Supplemental Material. Notice that $Q^{(2)}$ 
corresponds to a symmetric channel, and hence 
the maximal mutual information is given by the analytical expression 
$I^{(2)}=\log 3 -H(\{
\tilde \gamma, \tilde \gamma, 1-2 \tilde \gamma 
\})$, with $\tilde \gamma $ as in Eq. (A.12).  
The maximisation of the mutual information pertaining to
$Q^{(1)}$ can be obtained by the algorithm of Eq. (\ref{ari}), thus
giving $I^{(1)}$.  In Fig. \ref{fig:c3damp} we plot the detected capacity, corresponding to 
$C_{DET}=\max \{ I^{(1)},I^{(2)}\}$, versus the damping
parameters $\gamma _{01}$ and $\gamma _{02}$. 

The present example shows that our method allows to derive lower
bounds to the classical capacity of quantum channels which are even
theoretically poorly studied.  In addition, our method provides also the
explicit form of the encoding corresponding to the detected
lower bound. Before concluding, we notice that when
the measurement bases are badly matched with respect to the structure
of the unknown channel, then the detected capacity may give a bound to
the classical capacity which becomes looser. An example is provided in
the Supplemental Material, for the case of a dephasing channel for
qubits on an unknown basis. Moreover, we want to point out that there
are situations where a priori information is available on the form of
the channel, but quantum process tomography cannot be performed
because the number of allowed measurements is not enough. A simple
example is the case of a channel whose structure is known to be of the
form as in Eq. (\ref{em}), but only two measurements are available. On
the other hand, in the same scenario, our method provides a
significant detected capacity. A further 
example is given by a Pauli channel followed by a phase rotation,
which is explicitly reported in the Supplemental Material, along with a
quantitative analysis of the tradeoff between the available a priori
information on the probability distribution for the value of the
unknown phase and the performance of our detection method. 

In summary, we presented an efficient method to detect lower bounds to
the classical capacity of noisy quantum channels with few local
measurements, by testing orthogonal ensembles and output measurements,
along with classical optimisation algorithms.  The method can be
applied to completely unknown quantum channels and to all situations
where quantum process tomography is not available.  The scheme we
presented can be easily implemented in the lab with present day
technology, e.g. as in Ref. \cite{exp}.

\newpage

\widetext 

\setcounter{page}{1}
\setcounter{equation}{0}
\setcounter{figure}{0}
\appendix*
\section{SUPPLEMENTAL MATERIAL}
\subsection*{Capacity for classical binary asymmetric channels} 
Without loss of generality we can fix the labeling of zeros and ones
in a classical binary channel such that $0\leq \epsilon _0\leq \frac 12$,
$\epsilon _0 \leq \epsilon _1$, and $\epsilon _0 \leq 1 - \epsilon
_1$, where $\epsilon _0 $ denotes the error probability
of receiving $1$ for input $0$ 
and $\epsilon _1 $ denotes the error probability
of receiving $0$ for input $1$.    The Shannon capacity $C_B(\epsilon
_0, \epsilon _1)$ of the binary
channel is given by 
the maximum over the prior probability $p_0$ of the mutual information 
\begin{eqnarray}
I(p_0)=H[p_0 (1-\epsilon _0) + (1-p_0)
  \epsilon _1]- p_0 H[\epsilon _0] -(1-p_0) H[\epsilon _1]\;.\label{ipo}
\end{eqnarray}
From the condition $\frac {\partial I(p_0)}{\partial p_0}=0$ and
straightforward algebra one achieves
\begin{eqnarray}
p_0= \frac{1-\epsilon _1 (1+z)}{(1-\epsilon _0-\epsilon _1)(1+z)}\;,\label{po}
\end{eqnarray}
where 
\begin{eqnarray}
z=2^{\frac {H[\epsilon _0]-H[\epsilon _1]}{1-\epsilon _0 -\epsilon _1}}\;.
\end{eqnarray}
By substituting the optimal value (\ref{po}) of $p_0$ in
Eq. (\ref{ipo}) and simplifying, one obtains the
capacity
\begin{eqnarray}
C_B(\epsilon _0, \epsilon _1)=\log \left[ 1+ 
2^{\frac {H[\epsilon _0]-H[\epsilon _1]}{1-\epsilon _0 -\epsilon
  _1}}\right ] +\frac{\epsilon _0}{1-\epsilon _0 -\epsilon
  _1}H[\epsilon _1] -\frac{1-\epsilon _1}{1-\epsilon _0 -\epsilon
  _1}H[\epsilon _0]
\;.\label{c01} 
\end{eqnarray}
In the limiting case $\epsilon _0=\epsilon _1 =\epsilon $ one
recovers the classical capacity for the binary symmetric channel 
\begin{eqnarray}
C_B(\epsilon , \epsilon )= 1-H[\epsilon ]\;.
\end{eqnarray}
For $\epsilon _0=0$, only input 1 is affected by error, and 
one obtains the capacity of the so-called
$Z$-channel
\begin{eqnarray}
C_B(0,\epsilon )=\log [1 + (1-\epsilon ) \epsilon ^{\frac {\epsilon
    }{1-\epsilon }}]\;.
\end{eqnarray}

\subsection*{Detected capacity for a stretched damping channel}
A stretched damping channel can be specified by the parameters $\lambda _3=1-\gamma $,
$t_3=\gamma $, and $\lambda _1=\lambda _2=s$, with $\gamma \in [0,1]$
and $ |s|\leq
\sqrt{1-\gamma }$.  Applying Eq. (11) of the main text, one obtains 
\begin{eqnarray}
C_{DET}(\gamma ,s) \!=\!\max
\Big \{  
1-H\Big (\frac{1-|s|}{2}\Big ),C_B
(0,\gamma ) \Big \}\,.
\end{eqnarray}
 
\begin{figure}[htb]
\includegraphics[scale=0.6]{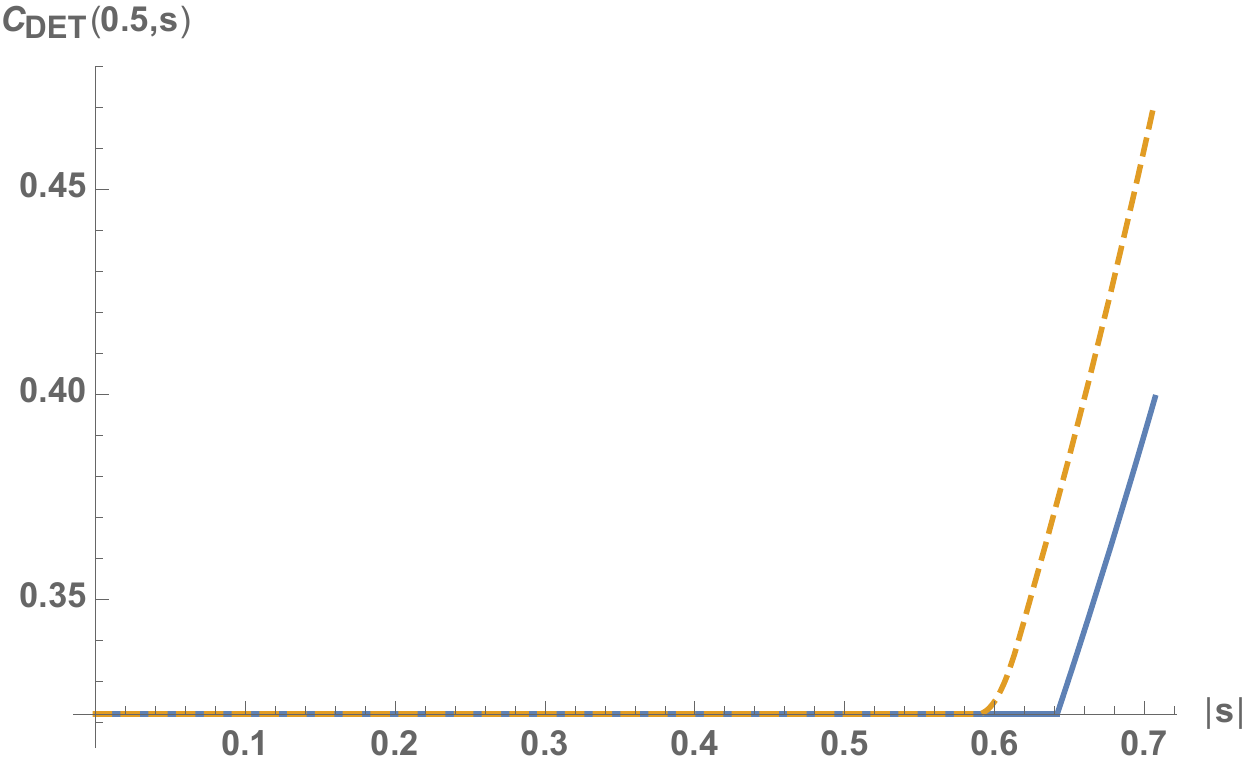} 
  \caption{Detected classical capacity $C_{DET}(\gamma ,s)$ for a
    stretched damping channel with damping parameter $\gamma =0.5 $,  
    versus allowed values of
    stretching $s$ (solid line), along with the 
    Holevo capacity (dotted line).}
  \label{fig:det2}
\end{figure} 
In Fig. 1. we plot the detected capacity for 
damping parameter $\gamma =\frac 12$,  versus the respective allowed values
of the stretching parameter $s$, along with the Holevo capacity $C_1$ 
(obtained numerically). In this case $T(1/2, 1/2)=
\frac{\ln  2}{2}$, and then from Eqs. (15) and (16) of the main text
it follows that 
the channel is pseudoclassical for $|s|\leq \sqrt \frac{\ln
  2}{2}\simeq 0.589$, for which $C_{DET}=C_1=C_B(0,1/2)=\log \frac54
\simeq 0.32193 $.

\subsection*{Extremal qubit channels} 

We consider here extremal qubit channels, which, up to rotations, 
can be parametrized as 
$\lambda _1= \cos \alpha $, $\lambda _2 =\cos
\beta $, $\lambda _3=\lambda _1 \lambda _2 =\cos \alpha \cos \beta $,
and $t_3=\sin \alpha \sin \beta $, with $0\leq \alpha \leq \beta \leq
\frac \pi 2$. It follows that $\epsilon _0^{x}=\epsilon _1^{x}=
\sin^2 \frac \alpha 2$, 
$\epsilon _0^{y}=\epsilon _1^{x}=\sin^2 \frac \beta 2$, $\epsilon _0^{z}=\sin^2 \frac
{\beta -\alpha} {2}$, and  $\epsilon _1^{z}=\sin^2 \frac
{\beta +\alpha} {2}$.    
Since $H(x)$ is monotonically increasing in $x\in
      [0,\frac 12]$, one easily finds that the detected capacity is
      given by 
\begin{eqnarray}
C_{DET}(\alpha ,\beta )=\max \left \{ 
1-H \left (\sin ^2 \frac \beta 2\right ), C_B
\left (\sin^2 \frac{\beta -\alpha }{2}, \sin^2 \frac {\beta +\alpha
}{2}\right ) \right\}
\;.\label{cabb}
\end{eqnarray}
We numerically checked that the first term is always the maximizer in
Eq. (\ref{cabb}), for all values of $\alpha $ and $\beta $. In fact, 
numerically one also can check that 
these channels never satisfy the pseudoclassicality condition in
Eq. (15) of the main text, except for
the degenerate unital case of $\alpha =0$ or $\beta = \frac \pi 2$, 
otherwise one would have $C_1=C_{DET}=
C_B \left (\sin^2 \frac{\beta -\alpha }{2}, \sin^2 \frac {\beta +\alpha
}{2}\right ) $ for some values of $\alpha $ and $\beta $.

\subsection*{Transition matrices for a V-shaped qutrit channel}
\begin{figure}[h!]
\begin{picture}(120,70)(0,0)
\put(11,50){\line(1,0){44}}
\put(60,44){\line(1,0){44}}
\put(35,10){\line(1,0){44}}
\put(32,50){\vector(1,-2){20}}
\put(82,44){\vector(-1,-2){17}}
\put(3,50){\makebox(0,0){$\ket{2}$}}
\put(112,44){\makebox(0,0){$\ket{1}$}}
\put(87,10){\makebox(0,0){$\ket{0}$}}
\end{picture}
\caption{Decay processes for a three-level atom with V-shaped pattern}
\label{fig:3levelatom}
\end{figure}
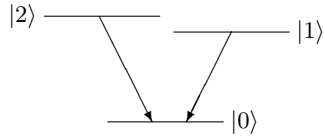

A decay process for a three-level system in $V$-shaped
configuration is depicted in Fig. \ref{fig:3levelatom}. 
The transition matrices $Q^{(1)}$ and $Q^{(2)}$ for the channel ${\cal
E}$ with Kraus operators as in Eq. (19) of the main text  
correspond to the conditional probabilities 
pertaining to the bases $B_1$ and $B_2$ in Eq. (20), namely 
\begin{eqnarray}
Q^{(i)}=p^{(i)}(m|n)={}_i \langle m | {\cal E} (|n \rangle _i {}_i
\langle n |) |m  \rangle _{i}  
\;.
\end{eqnarray}
A straightforward calculation gives 
\begin{eqnarray}
Q^{(1)}=
\begin{pmatrix} 1 & \gamma _{01} & \gamma _{02}\\ 
0 & 1- \gamma _{01} & 0 \\
0 & 0 & 1- \gamma _{02} 
\end{pmatrix},
\;
\end{eqnarray}
and 
\begin{eqnarray}
Q^{(2)}=
\begin{pmatrix} 1 -2 \tilde \gamma & \tilde \gamma & \tilde \gamma \\
\tilde \gamma &  1 -2 \tilde \gamma & \tilde \gamma  \\
\tilde \gamma & \tilde \gamma & 1 -2 \tilde \gamma 
\end{pmatrix}
\;,
\end{eqnarray}
where 
\begin{eqnarray}
\tilde \gamma = \frac13 -\frac 19 \left(\sqrt{1-\gamma _{01}}
+\sqrt{1-\gamma _{02}}+\sqrt{(1-\gamma _{01})(1-\gamma _{02})}\right )
\;.\label{gtil}
\end{eqnarray} 
Notice that $Q^{(2)}$ corresponds to a symmetric channel, i.e. all
rows (and columns) are permutation of each other. Hence, there
is no need of numerical optimisation over the prior probability $p^{(2)}_n$, since
the optimal is the uniform distribution.  The corresponding 
maximal mutual information is given by the analytical expression 
$I^{(2)}=\log 3 -H(\{
\tilde \gamma, \tilde \gamma, 1-2 \tilde \gamma 
\})$. The maximisation of the mutual information pertaining to
$Q^{(1)}$ can be obtained by the Blahut-Arimoto algorithm in
Eq. (6) of the main text, which
provides the optimal prior probability $p^{(1)}_n$ and the corresponding value of $I^{(1)}$.  
The detected capacity is then given by $C_{DET}=\max \{ I^{(1)}, I^{(2)} \}$.   

\subsection*{Detected capacity for qubit dephasing channels on an unknown basis} 

In the main text we noticed that when the measurement bases are badly
matched with respect to the structure of the unknown channel, then the
detected capacity may give a bound to the classical capacity which
becomes looser. This is the price to pay if one has to avoid complete
process tomography and the channel is completely unknown.   
We give here an example for the case of a dephasing
channel for qubits with probability $p$ on an unknown basis. This
channel can be written as 
\begin{eqnarray}
{\cal E}(\rho )= (1-p)\rho + p \,
\vec \sigma  \cdot \vec n \,
\rho \,\vec \sigma  \cdot \vec n \,
\;,\label{dro}
\end{eqnarray}
where $\vec\sigma $ is the vector of Pauli operators $(\sigma_x,\sigma_y,
\sigma_z)$, and $\vec n = (\sin \theta \cos \phi , \sin \theta
\sin \phi, \cos \theta)$ is a unit vector on the Bloch
sphere ($\theta \in [0,\pi/2]$ and $\phi \in [0,2\pi]$). 
The classical capacity is one bit and is clearly achieved by encoding on the
eigenstates of $\vec \sigma  \cdot \vec n $ which are noise-free.  

Consider now the detection method with measurements of the Pauli
matrices. The conditional probabilities of  outcomes $\pm 1$ for the measurement of
$\sigma _{i}$  (with $i=x,y,z$) at the output of the dephasing channel in
Eq. (\ref{dro})  with input state 
$\rho ^{(i)}_{\mp}=\frac 12 (1 \mp \sigma _i)$ 
are given by  
\begin{eqnarray}
&&p^{(x)}(\pm 1 | \mp 1)= p (\cos^2\theta +\sin^2 \theta \sin^2 \phi)\;,
  \nonumber \\
&&p^{(y)}(\pm 1 | \mp 1)= p (\cos^2\theta +\sin^2 \theta \cos^2 \phi)\;,
  \nonumber \\& &
 p^{(z)}(\pm 1 | \mp 1)= p \sin^2 \theta \;.
\end{eqnarray}
Hence, these three conditional probabilities correspond to three
classical binary symmetric channels, and the detected capacity is given by 
\begin{eqnarray}
C_{DET}(p)=1- \min \{H[p (\cos^2\theta +\sin^2 \theta \sin^2 \phi)],
H[p (\cos^2\theta +\sin^2 \theta \cos^2 \phi)],
 H[p \sin^2 \theta ] \}\;.\label{csn}
\end{eqnarray}
Notice that the detected capacity in Eq. (\ref{csn}) is invariant 
for $\phi \rightarrow \phi +\frac \pi 2$. The effect of basis-mismatch
can be detrimental, and in the present example, the
worst-case scenario corresponds to $\phi = \pi/4 $ and $\theta =\mbox{arcos}
(1/\sqrt{3})$, for which $C_{DET}(p)=1-H(2p/3)$. 
\begin{figure}[htb]
  \includegraphics[scale=0.6]{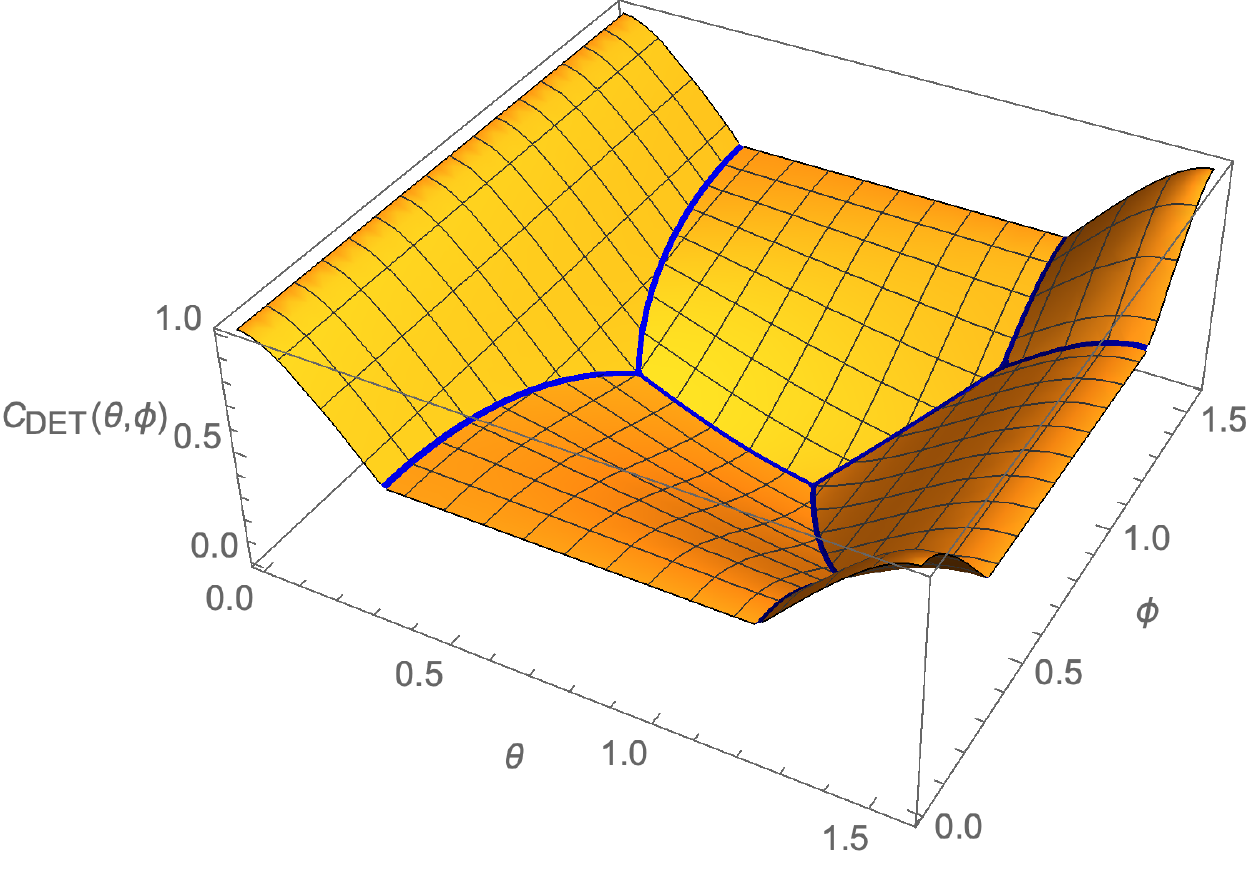}
  \caption{Detected classical capacity for a qubit dephasing  
channel on unknown basis with error probability $p=0.9$ versus basis
parameters $\theta ,\phi $.}
  \label{fig:depo}
\end{figure} 
In Fig. 3 we plot the detected capacity for $p=0.9$ versus the angles
$\theta $ and $\phi $. 
  
\subsection*{Detected capacity for a Pauli channel followed by a phase
  rotation}
Let us consider a Pauli channel with unknown probabilities
$p_x$, $p_y$, and $p_z$, followed by a rotation around the $\sigma _z$-axis by an
unknown angle $\phi $, i.e. 
\begin{eqnarray}
{\cal E}(\rho )=e^{i \frac \phi 2 \sigma _z}
[(1-p_x-p_y-p_z)\rho +
p_x \sigma _x \rho \sigma _x + p_y \sigma _y \rho \sigma _y +p_z
\sigma _z \rho \sigma _z]e^{-i \frac \phi 2 \sigma _z}
\;. \label{phir}
\end{eqnarray}
One can easily verify that our detection method based just on the
measurements of $\sigma _x$, $\sigma _y$, and $\sigma _z$ provides a
significant detected capacity. 
The conditional probabilities of outcomes $\pm 1$ for the measurement of
$\sigma _{i}$  (with $i=x,y,z$) at the output of the channel with input state 
$\rho ^{(i)}_{\mp}=\frac 12 (1 \mp \sigma _i)$ 
are given by  
\begin{eqnarray}
&&p^{(x)}(\pm 1 | \mp 1)= \frac{1-\cos \phi}{2} +(p_y+p_z)\cos \phi \;,
  \nonumber \\
&&p^{(y)}(\pm 1 | \mp 1)= \frac{1-\cos \phi}{2} + (p_x+p_z)\cos \phi 
\;,
  \nonumber \\& &
 p^{(z)}(\pm 1 | \mp 1)= p_x+p_y
\;,\label{pxyz}
\end{eqnarray}
and hence Eq. (12) of the main text is replaced by 
\begin{eqnarray}
C_{DET}(\phi) =1- \min\left \{H\left [\frac{1-\cos \phi}{2} +
  (p_y+p_z)\cos \phi \right ],
H\left [\frac{1-\cos \phi}{2} + (p_x+p_z)\cos \phi \right ],
H[p_x+p_y]\right \} 
\;.\label{cph}
\end{eqnarray}
We notice that even if the structure of
the channel were known, the measurement results would not 
allow to perform quantum process tomography, since the
expectations $\Tr [\sigma _x {\cal E}(\sigma_y )]$ and $\Tr [\sigma _y
  {\cal E}(\sigma_x )]$ are not available. This can also be recognized by
the fact that Eqs. (\ref{pxyz}) obtained by the measurements cannot be
solved to determine the four variables $p_x,p_y,p_z$, and $\phi $. 
 
\par When some prior information is available about the quantum
channel, one
can also quantify a sort of tradeoff between such information 
and the performance of our detection method. For example, let
us formalize our uncertainty about $\phi \in [-\pi ,\pi ]$ by 
a von Mises probability density 
\begin{eqnarray}
p(\phi)= \frac{\exp (K_\phi \cos \phi )}{2 \pi  I_0 (K_\phi)}
\;,\label{von2}
\end{eqnarray}
where $I_0(x)$ denotes the modified $0$-order Bessel function, and
$K_\phi$ is a parameter which measures the
concentration (i.e., it is analogous of the reciprocal of variance
for normal distributions). In other words, Eq. (\ref{von2})
encapsulates our knowledge that the most likely  value of $\phi $ is
zero, with increasing confidence for increasing
values of $K_\phi $  (clearly, if the expected value of $\phi $ is
different from zero, say $\phi^*$, one could consequently change the
set of measured observales to $\{\sigma _x \cos \phi ^* 
+ \sigma _y \sin \phi^* , \sigma _y \cos \phi ^*  -\sigma _x \sin \phi ^* , \sigma
_z\}$ and achieve the same performance).  

The expected detected capacity can then be evaluated by the weighted
average of Eq. (\ref{cph}) with the function (\ref{von2}), namely 
\begin{eqnarray}
C_{DET}=\int _{-\pi} ^\pi C_{DET }(\phi ) p(\phi ) d \phi\;. \label{}
\end{eqnarray}
In Fig. 4 we plot a specific result for channel parameters $p_x=0.15$,  
$p_y=0.05$,  and $p_z=0.1$,  versus the concentration parameter
$K_\phi$. The average detected capacity grows from  $C_{DET}\simeq
0.3031 $ (corresponding to 
$K_\phi =0$, i.e. a flat distribution and hence total ignorance on
the phase $\phi $) to the theoretical classical
capacity given by $C=1-H(0.15) \simeq 0.3902$ for increasing values of
$K_\phi$, i.e. of the available information on $\phi $. 

\begin{figure}[htb]
\includegraphics[scale=0.6]{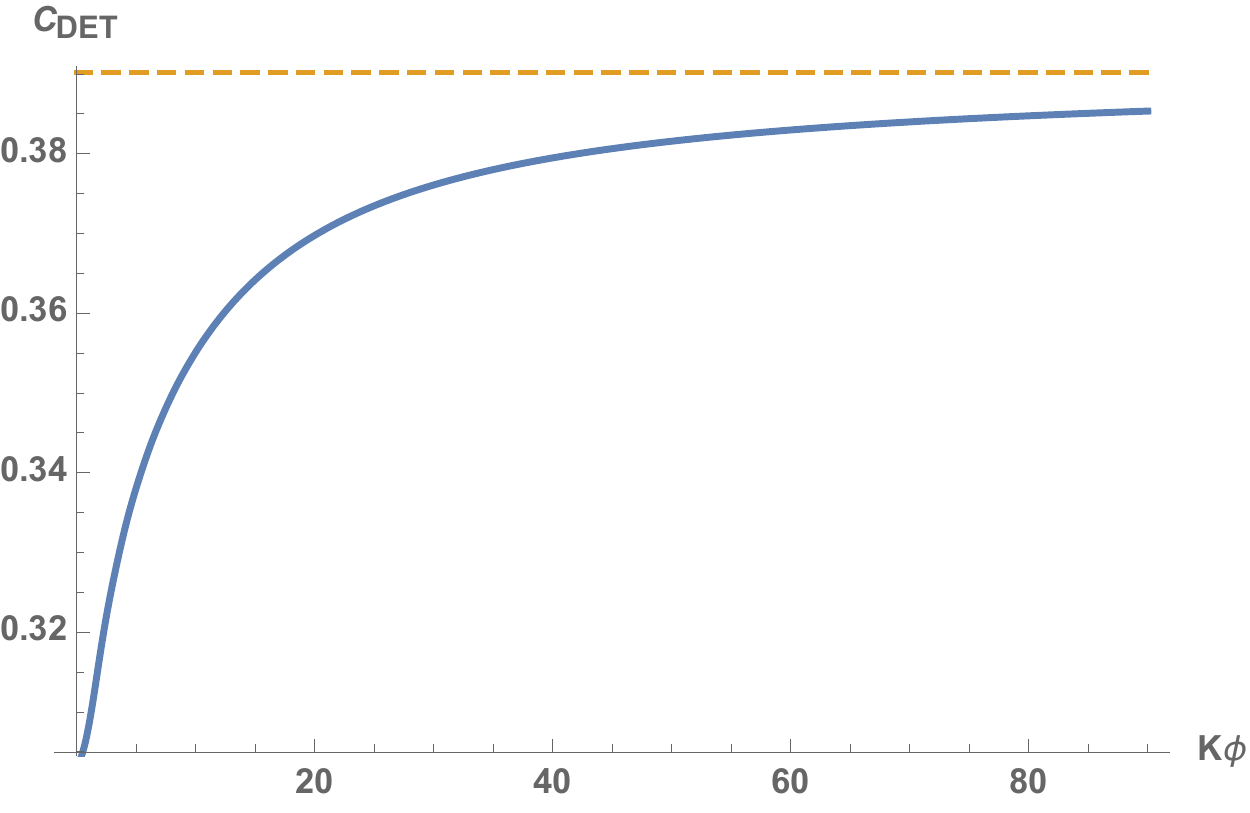}
\caption{Average detected classical capacity for a Pauli channel
    with error propabilities $p_x=0.15$,  
$p_y=0.05$,  and $p_z=0.1$, followed by a $\sigma _z$-phase rotation by
    $\phi $,  versus the concentration parameter $K_\phi$ 
characterizing the available prior
information on $\phi $ described by the probability density in
Eq. (\ref{von2}). The detected capacity grows from
 $C_{DET}\simeq 0.3031$ for $K_\phi =0$ (complete ignorance on $\phi$)
 towards the theoretical classical capacity $C=1-H(0.15) \simeq
 0.3902$ for increasing values of $K_\phi$, 
i.e. of the available information on $\phi $.} 
  \label{fig:depo}
\end{figure}

\end{document}